\documentstyle[aps,multicol, psfig,epsfig]{revtex}
\renewcommand{\narrowtext}{\begin{multicols}{2} \global\columnwidth20.5pc}
\renewcommand{\widetext}{\end{multicols} \global\columnwidth42.5pc}

\def\nb #1{{\hbox{\bf #1}}}

\def\bOmega{\overline{\Omega}}

\def\emn{\varepsilon_{\mu\nu}}
\def\Xl{x_1}        
\def\Yl{y_1}        
\def\Xr{x_2}        
\def\Yr{y_2}        
\def\Xlinit{x_1^{(0)}}       
\def\Ylinit{y_1^{(0)}}       
\def\Xrinit{x_2^{(0)}}       
\def\Yrinit{y_2^{(0)}}       
\def\Xi{x_i}

\def\Vl{v_1}        
\def\Vr{v_2}        
\def\Ll{L_1}        
\def\Lr{L_2}        

\newcommand{\gsim}{\buildrel{>}\over{\sim}}

\begin{document}


\title{Scattering of vortex pairs in 2D easy-plane ferromagnets}
\author{A.S. Kovalev}
\address{Institute for Low Temperature Physics and Engineering,
47 Lenin Ave., 61164 Kharkov, Ukraine}
\author{S. Komineas and F.G. Mertens}
\address{Physikalisches Institut, Universit\"at Bayreuth, D-95440 Bayreuth,
Germany}

\date{\today}
\maketitle

\begin{abstract}
Vortex-antivortex pairs in 2D easy-plane ferromagnets have characteristics
of solitons in two dimensions.
We investigate numerically and analytically the dynamics 
of such vortex pairs. In particular
we simulate numerically the head-on collision of two pairs 
with different velocities for a
wide range of the total linear momentum of the system.
If the momentum difference of the two pairs is small, the vortices 
exchange partners, scatter at an angle depending on this difference,
and form two new identical pairs.
If it is large, the pairs pass through each other without 
losing their identity.
We also study head-tail collisions.
Two identical pairs moving in the same direction are bound into a moving
quadrupole in which the two vortices as well as the two antivortices
rotate around each other.
We study the scattering processes also analytically in the
frame of a collective variable theory,
where the equations of motion for a system of four vortices constitute
an integrable system. The features of the different collision
scenarios are fully reproduced by the theory.
We finally compare some aspects of the present soliton scattering
with the corresponding situation in one dimension.
\end{abstract}

\pacs{}

\narrowtext

\section{ Introduction}

The statics and dynamics of magnetic vortices is already 
an old subject \cite{kovalev79,kosevich83,nikiforov,huber}. 
An increasing interest
in the problem has arisen again \cite{gouvea,mertens1,mertens2}
which is connected with the synthesis and experimental study
of new low-dimensional magnetic compounds such as two-dimensional 
magnetic lipid layers, organic intercalated quasi-2D layered magnets
and HTSC-materials in the antiferromagnetic state.

Magnetic vortices play an important role in easy-plane magnets.
They are the main ingredients in the Kosterlitz-Thouless phase transition.
At a finite temperature the density of vortices
is large and they should give a considerable contribution to the 
dynamic correlations.

We already have a detailed picture of the dynamics in a system with
a small number of vortices.
An isolated vortex in an infinite system can only move together with
the background flux \cite{nikiforov,paptom}. 
Two vortices interact and undergo Kelvin motion if they have opposite
topological charges while they move one around the other when they have 
the same charge. We shall call the pair of two vortices with 
opposite topological charge a vortex-antivortex pair (V-A pair).
Such pairs can have the characteristics of a soliton in the sense that they
move coherently with some constant velocity.
Solitons of this kind have been numerically investigated in some magnetic
systems \cite{semi,cooper}.
Analogous V-A pairs have been studied
in superfluids \cite{pitaevskii,jones1,jones2}, nonlinear optics 
\cite{luther} and hydrodynamics \cite{lamb}.

In a system with a lot of vortices the picture becomes 
accordingly more complicated.
If we suppose a dilute vortex gas, the average
velocity is $\overline{V} \sim \sqrt{\rho} \sim 1/L$, where $\rho$ is the
density of vortices and $L$ the average distance between them.
The interaction energy is proportional to the logarithm of the average
distance. This picture
should be realistic above the Kosterlitz-Thouless temperature.
At low temperatures vortex-antivortex pairs are expected to form.
The energy of a pair is finite and it is proportional to the logarithm
of its size.
The interaction potential between them is
inversely proportional to the second power of their size.
One may be tempted to treat the vortex pairs as elementary weakly interacting
particles. However, 
their dynamics is not Newtonian and most importantly they have 
an internal structure which may change during interaction.

In a dense enough gas of vortices, interactions among traveling V-A pairs
are unavoidable.
In particular, any change in the number of vortices present in the system,
or the number of vortices in equilibrium, should be a direct or
indirect result of the scattering among V-A pairs.
Our article is devoted to head-on and head-tail collisions between
V-A pairs in easy-axis ferromagnets, i.e. to the case of zero total
angular momentum of the system.
Our study is both numerical and analytical. A collective coordinate theory
is found to be particularly successful and provides the basis for
a clear picture of the dynamics.

Our results should also be relevant for a variety of other systems
where V-A pairs have been found.
Furthermore comparisons can be made to the well-studied 
soliton interactions in one space dimension.

The outline of the rest of the paper is as follows.
In Sec. II we give a short description of the system and an account
of the dynamics of vortices and vortex-antivortex pairs.
In Sec. III we present numerical simulations for collisions
between V-A pairs. Sec. IV presents a theory which explains the
features of the dynamical behavior of vortex pairs.
Our concluding remarks are contained in the last Sec. V.

\section{ Vortices and vortex-antivortex pairs}

We consider the classical two-dimensional Heisenberg ferromagnet
with a uniaxial anisotropy of the easy-plane type.
The corresponding Hamiltonian has the form
  \begin{equation}
  \label{eq:heisenberg}
{\cal H}=-J\sum_{(n,m)}\left( {\bf S}_n \cdot {\bf S}_m \right)
 + {\beta \over 2} \sum_{n} (S^z_n)^2,
  \end{equation}
where ${\bf S}_n$ denotes the spin variable at site $n$,
$S^z_n$ is the third component of ${\bf S}_n$.
The first summation runs over the nearest-neighbor pairs.
The exchange constant $J$ and the single-ion anisotropy constant $\beta$
are positive.
We treat the spin ${\bf S}$ as
a classical vector of constant length.
Usually the magnetic anisotropy is small: 
$\beta/J\! \sim\! 10^{-2}$ or $10^{-3}$. In this case we can use
a continuum approximation of the Hamiltonian.
We define two fields $m\! =\! S^z$ and $\Phi\!=\!\arctan(S^y/S^x)$,
where $S^x, S^y, S^z$ are the cartesian components of the spin.
In terms of these variables the continuum version of the Hamiltonian reads
  \begin{equation}
  \label{eq:hamiltonian}
{\cal H}={\beta \over 2}\,\int dx dy\, \Bigl[\,{(\nabla m)^2 \over 1-m^2}+
(1-m^2)\,(\nabla \Phi)^2+m^2\,\Bigr].
  \end{equation}
The coordinates $x,y$ are measured in  units of the ``magnetic length'' 
$l_0=\sqrt{J/\beta}$.
$m, \Phi$ are canonically conjugate fields and
the equations of motion have the Hamiltonian form \cite{kosevich90}
  \begin{equation}
  \label{eq:canonical}
{\partial \Phi \over \partial t} = {\delta {\cal H} \over \delta m}, \qquad
{\partial m \over \partial t} = -{\delta {\cal H} \over \delta \Phi}\;,
  \end{equation}
explicitely
  \begin{eqnarray}
  \label{eq:m-phi}
\frac{\partial \Phi}{\partial t} & = &  m -\frac{\Delta m}{1-m^2}-
\frac{m\, (\nabla m)^2}{(1-m^2)^2}-m\,(\nabla \Phi)^2\,, \nonumber \\
\noalign{\medskip}
\frac{\partial m}{\partial t} & = & (1-m^2)\,\Delta \Phi -2\,m\,
\nabla m \nabla \Phi \,.
  \end{eqnarray}

Our numerical algorithm uses the formulation
through the stereographic variable
  \begin{equation}
\label{eq:omega}
\Omega=\sqrt{\frac{1-m}{1+m}}\;\exp(i\Phi)\;.
  \end{equation}
This satisfies the equation
  \begin{equation}
\label{eq:omegaeq}
i\;\frac{\partial \Omega}{\partial t}=-\Delta \Omega +
{2\,\bOmega \over 1+\Omega  \bOmega}\; {\partial}_{\mu}\Omega\,
{\partial}_{\mu}\Omega- 
{1-\Omega  \bOmega \over 1+ \Omega  \bOmega}\; \Omega \,,
  \end{equation}
where $\bOmega$ denotes the complex conjugate of $\Omega$.
This variable was used in the solution
of the Landau-Lifshitz equation in one dimension
since, in terms of it, the soliton solutions attain
their simplest form \cite{kosevich90,bogdan}.

In studying statics and dynamics for the above model,
the topological density 
  \begin{equation}
\label{eq:vorticity}
\gamma=\frac{\partial m}{\partial x}\frac{\partial \Phi}{\partial y}
-\frac{\partial m}{\partial y}\frac{\partial \Phi}{\partial x}\,
  \end{equation}
is a most useful quantity.
It has been called the "local vorticity" since it plays here a role
analogous to the ordinary vorticity in fluid dynamics \cite{paptom,afm}.
The integrated topological density 
  \begin{equation}
\label{eq:totalvorticity}
\Gamma = \int{\gamma\, dxdy} 
  \end{equation}
is an invariant and takes values which are integral multiples of $2\pi$
for the vortex solutions that we shall discuss here.

The vorticity density enters
the definitions of the linear and angular momentum of the theory
which read \cite{paptom}
  \begin{equation}
\label{eq:linearmom}
P_\mu = -\emn \int{x_\nu\, \gamma\, dxdy}, \quad \mu,\nu=1,2\,,
  \end{equation}
  \begin{equation}
\label{eq:angularmom}
 \ell = \int{(x^2+y^2)\, \gamma\, dxdy}.
  \end{equation}

The role of the total vorticity in the dynamics can be also
appreciated through the definition of the so-called
"gyrocoupling vector" \cite{thiele} 
\begin{equation}
\label{eq:gyrovector}
{\bf G} = -\hat{\nb z}\,\Gamma\;,
\end{equation}
where $\hat{\nb z}$ is the unit vector in the third direction.
The gyrocoupling vector enters the equations 
which describe the dynamics of vortices
in a collective coordinate theory.

We now turn our attention to the discussion of topological excitations.
We take as a boundary condition that the field $\Phi$
is proportional to the polar angle $\phi$ at spatial infinity. 
We then obtain vortex solutions which have the form
\cite{kosevich83,nikiforov,gouvea}
  \begin{equation}
\label{eq:vortexphi}
\Phi=\kappa\;\arctan {y-Y \over x-X}\,,
  \end{equation}
  \begin{equation}
\label{eq:vortexm}
m = f(|{\bf r - R}|),
  \end{equation}
where $\kappa = \pm 1, \pm 2,...$ will be called the vortex number
and $(X,Y)$ is the position of the vortex center.
In the following we call the vortices with $\kappa < 0$ antivortices.
The magnetization field $m$ for a vortex can be found numerically
and has the following asymptotic behavior \cite{gouvea}:
  \begin{eqnarray}
\label{eq:masymptotic}
m & = & \lambda\,[1\;-\;a\,r^2],\quad r \rightarrow 0\,, \nonumber \\
\noalign{\medskip}
m & = & \lambda\,b\,\;\exp (-r)/\sqrt{r},\quad r \gg 1\,,
  \end{eqnarray}
where $r$ is the distance from the vortex center, $a,b$ are constants
and $\lambda\!=\!m(r\!=\!0)\!=\!\pm 1$ we call the "polarity" of the vortex.
The radius of the vortex is unity in our units.
We use in our numerical simulations only vortices and antivortices
with $\kappa \!=\! \pm 1$ and polarity $\lambda\!=\! 1$.

The total vorticity of the vortices (\ref{eq:vortexphi},\ref{eq:vortexm})
is $\Gamma\!=\!- 2\pi \kappa \lambda$.
The structure of the magnetic vortices (\ref{eq:vortexphi},\ref{eq:vortexm})
is similar to that of the vortices in a non-ideal Bose gas \cite{pitaevskii}
where the quantity $(1-m)$ is the density of the Bose particles.
However, the ferromagnetic vortices differ in that they
come with two possible values of the polarity $\lambda$.

The most impressive characteristic of the dynamics of an isolated vortex is
that it is spontaneously pinned in an infinite
medium. One can trace the reasons of this dynamical behavior
to their topological complexity which is reflected in the nonzero
value of $\Gamma$ \cite{paptom}. 
On the other hand, a vortex-antivortex pair undergoes Kelvin motion.
This motion
was studied in \cite{volkel91,volkel94} (see also \cite{papzakr})
for a large vortex separation.
In general, Kelvin motion sets in when the two vortices
have opposite total vorticities: $\kappa_1 \lambda_1 = -\kappa_2 \lambda_2$.

One can argue that
the simplest topologically nontrivial objects which can be
found in free translational motion should have
the form of a vortex-antivortex pair with $\Gamma\!=\!0$.
A conclusive numerical and analytical study
in an easy-plane ferromagnet was given in \cite{semi}
where the profiles of coherently moving structures were numerically calculated.
There is a branch of solitons with velocities varying from
zero to unity, which is the velocity of spin waves in the medium in our units.
For small velocities the solitons have indeed the form of a V-A pair
with a large separation $L$ between the vortex and the antivortex.
Their velocity is inversely proportional to the distance between them
\begin{equation}
v = {1 \over L}, \qquad L\gg 1.
\label{eq:1overl}
\end{equation}
When the velocity approaches that of spin waves in the medium the
vortex-antivortex character of the soliton is lost. The transition
occurs at $v \simeq 0.78$ in the sense that above this
velocity the spin does not reach the north pole at any point.

Similar ideas have a long history in hydrodynamics  where V-A pairs 
have been studied theoretically and experimentally
in 2D flows. It is known that a V-A pair undergoes Kelvin motion 
in a direction perpendicular
to the line connecting the centers of the vortex and the antivortex
and the velocity is inversely proportional to the distance between them
\cite{lamb,lugt}.
However, all the studies were made in the limit of point vortices,
that is all moving objects had a clear vortex-antivortex character.
We shall go beyond this situation in the present paper.

\section{Collisions of Vortex-Antivortex pairs: Numerical simulations}

The investigation of the interactions among the traveling V-A pairs 
comes as a natural next step after the well established theories
of the previous section.
In the following, we approach the subject through numerical simulations.
We simulate collisions of pairs
which are initially moving along the same line, say the horizontal $x$-axis.
We are interested both in head-on collisions, where the pairs
move initially in opposite directions and in head-tail collisions,
that is when they move in the same direction.

In the first set of our simulations, 
we confine ourselves to V-A pairs with small velocities
so that the topological characteristics remain distinct and 
the vortices in the pair retain their identity. In particular,
the distance
between the vortex and antivortex centers is larger than the size of
a single vortex ($L\! >\! 1$ in our conventions).
When the sizes of the simulated pairs
are large, that is the distance between the vortex and the antivortex
is large ($L \gg 1$),
we use as an initial condition the ansatz
\begin{equation}
\label{eq:ansatz4}
\Omega = \prod_{i=1}^4 \Omega_i\,, \qquad \Omega_i\;=\;
\sqrt{\frac{1-f_i}{1+f_i}}\;e^{i\Phi_i}\;,
\end{equation}
where $f_i$ and $\Phi_i$ are the functions
in Eqs.~(\ref{eq:vortexphi},\ref{eq:vortexm}). They represent vortices
which are centered at ${\bf R_i}\!=\!(X_i,Y_i)$ and have vortex numbers 
$\kappa_i$.
In particular we choose 
$\kappa_1\!=\!-\kappa_2,\;\; X_1\!=\!X_2\!\equiv\! \Xl,\;\; 
Y_1\!=\!-Y_2\!\equiv\! \Yl\;$
for the first pair which has a size $\Ll\!=\!2 \Yl$. We take accordingly
$\kappa_3\!=\!-\kappa_4,$ and $X_3\!=\!X_4\!\equiv\! \Xr,\;\;
Y_3\!=\!-Y_4\!\equiv\! \Yr\;$
for the second pair which has a size $\Lr\!=\!2 \Yr$ 
(we suppose $\Yl,\Yr\!>\!0$).
The above ansatz represents two V-A
pairs which are moving on the $x$-axis and are set in a collision course,
as long as the size of each pair is smaller than the distance between them:
$\Ll,\,\Lr \ll \delta \equiv |\Xl-\Xr|$.
As an alternative to the ansatz (\ref{eq:ansatz4}) we also use
the product ansatz of two V-A pair solitons of ref. \cite{semi}. 
This ansatz resembles in its gross
features the ansatz (\ref{eq:ansatz4}).

We perform the numerical simulations on a $500 \times 500$ mesh
with a uniform lattice spacing, typically $h=0.2$.
The time integration is performed by a fourth order
Runge-Kutta routine.
 
In \cite{rightangle} one of the present authors has investigated the
interaction process of two identical V-A pairs which collide head-on.
It was found that during the collision the vortices exchange their
partners and two new pairs are formed which are scattered at right angles.
The trajectories of the magnetic vortices are similar to
those in collisions of V-A pairs in hydrodynamics \cite{lamb,lugt} 
and they conform to a good accuracy to the formula 
$1/X_i^2+1/Y_i^2 = \hbox{const.}$
obtained in the 19th century \cite{grobli,greenhill}.

\begin{figure}
   \begin{center}
   \psfig{file=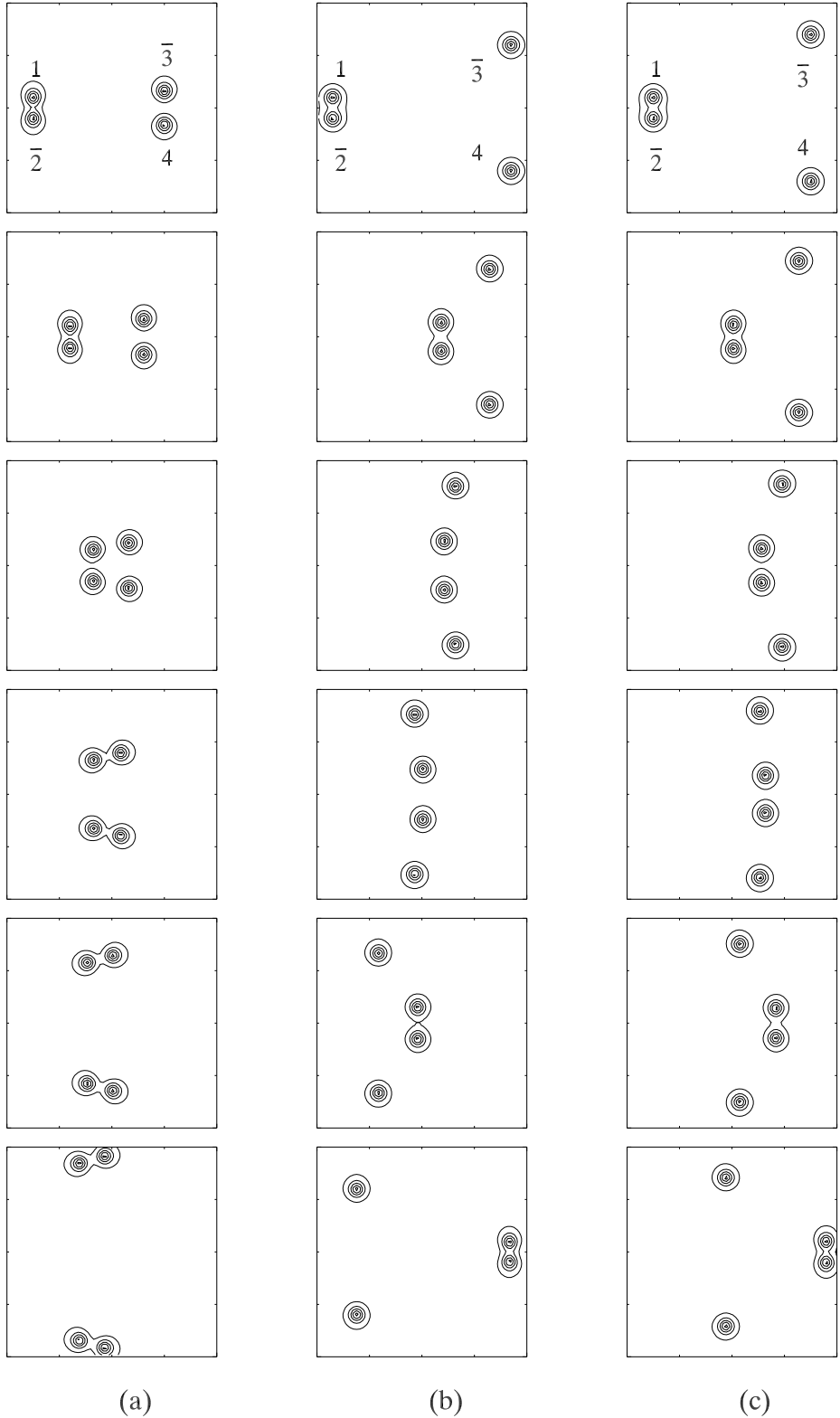,height=21.0cm,bbllx=50bp,bblly=130bp,bburx=450bp,bbury=760bp}
   \end{center}
   \caption{Head-on collisions between vortex-antivortex pairs.
We plot here contours of the third component of the spin
using the levels 0.1,0.3,0.5,0.7,0.9.
The numbers 1,4 denote the vortices and the $\overline{2},\overline{3}$
the antivortices.
In (a) we have the case
of a small difference between the linear momentum of the two pairs
and the vortices exchange partners and scatter at an angle.
In (b) the difference in momentum is larger. The pairs exchange partners,
follow a looping orbit and finally rejoin the initial partners and travel
along the initial direction of motion.
In (c) the momentum difference is large. The
two pairs pass through each other.
Note that the boxes presented here have dimensions $40\times 40$ while
the simulations were done in a space $100\times 100$.
   }
   \label{fig:head-on}
\end{figure}

\begin{figure}
   \begin{center}
   \psfig{file=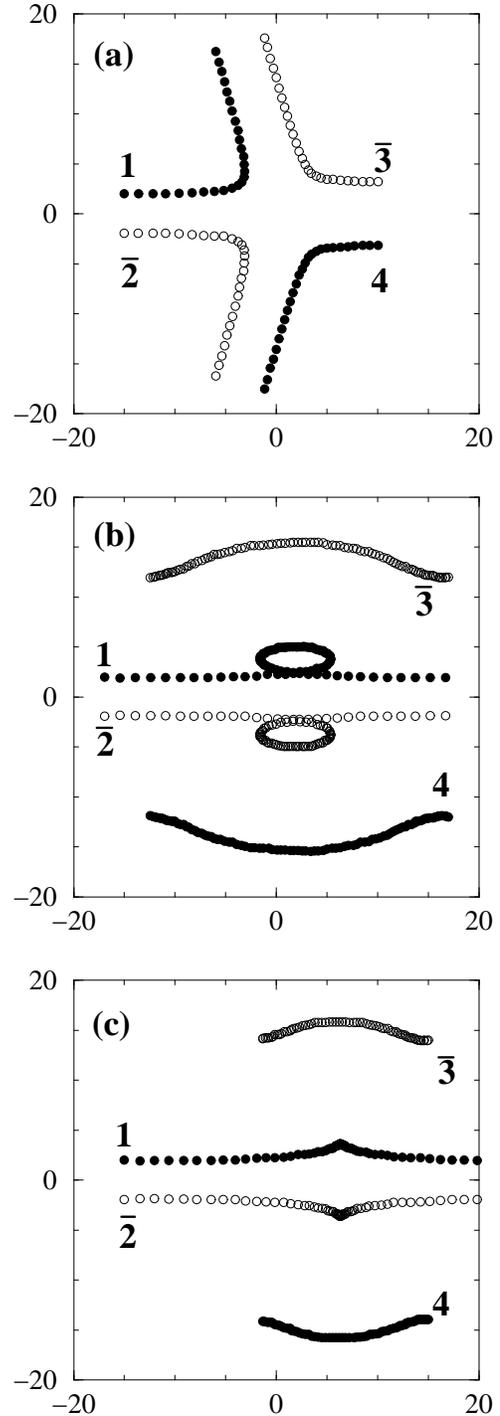,width=6.7cm,bbllx=120bp,bblly=40bp,bburx=380bp,bbury=765bp}
   \end{center}
   \caption{The orbits of the vortices and antivortices of
Fig.~\ref{fig:head-on} during their head-on collision.
The filled circles denote the position of vortices and the open
circles the position of the antivortices at successive and equal
times intervals. The numbers $1,\overline{2},\overline{3},4$ denote the initial
position of the vortices and antivortices.
   }
   \label{fig:orbit1}
\end{figure}

\begin{figure}
   \begin{center}
   \psfig{file=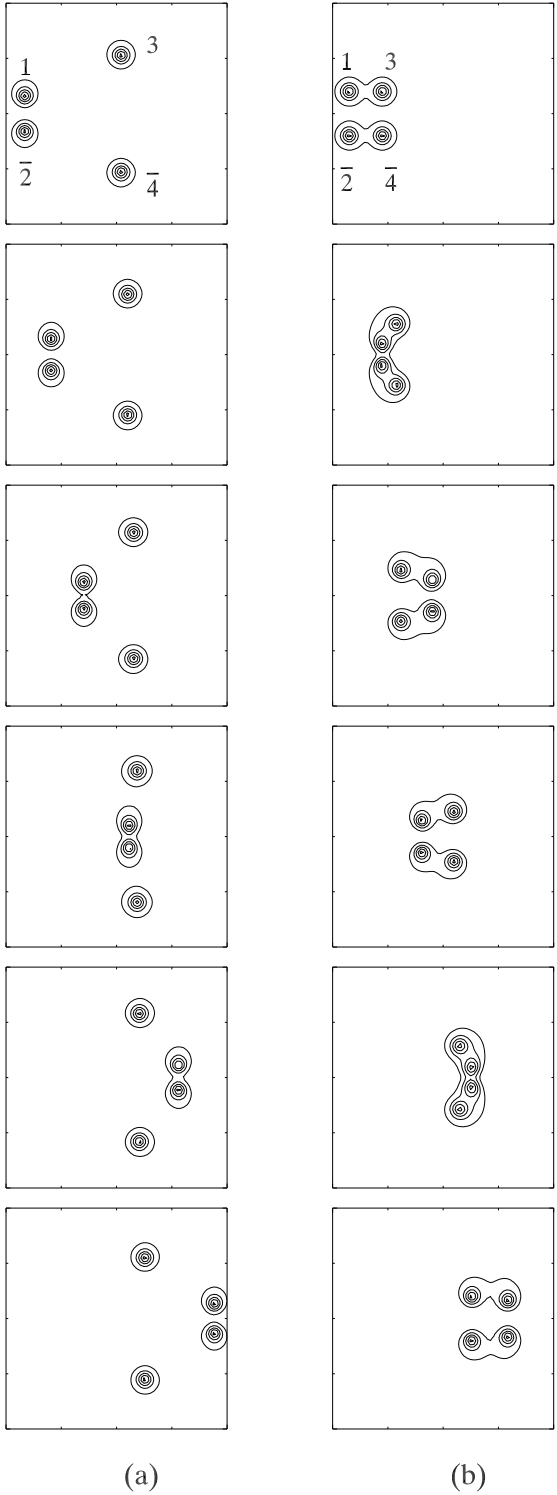,height=20.0cm,bbllx=50bp,bblly=130bp,bburx=310bp,bbury=760bp}
   \end{center}
   \caption{Head-tail collisions between vortex-antivortex pairs.
We plot here contours of the third component of the spin
using the levels 0.1,0.3,0.5,0.7,0.9.
The numbers 1,3 denote the vortices and the $\overline{2},\overline{4}$
the antivortices.
In (a) the momentum difference is large and the pairs pass through each other.
In (b) we have a propagating quadrupole state.
The boxes have dimensions $40\times 40$,
the simulations were done in a space $100\times100$.
   }
   \label{fig:head-tail}
\end{figure}

\begin{figure}
   \begin{center}
   \epsfig{file=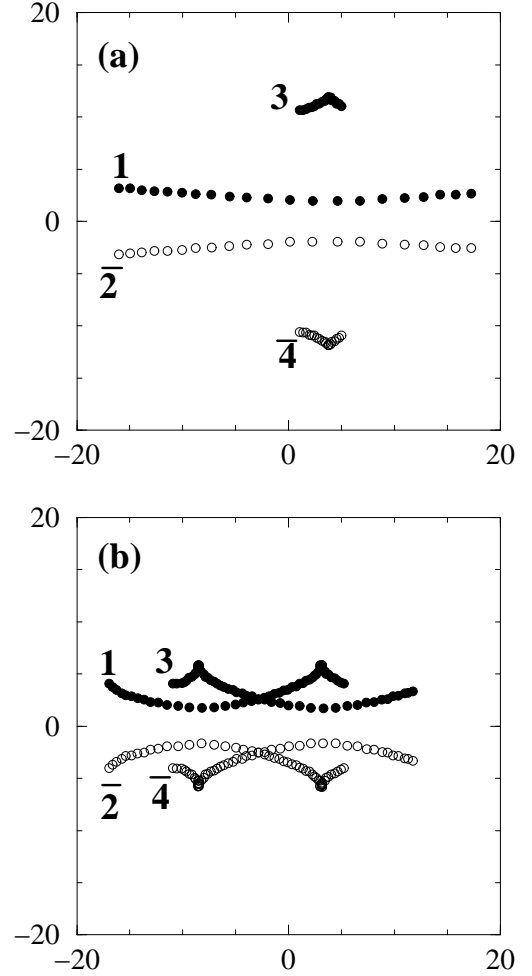,width=7.0cm,bbllx=120bp,bblly=190bp,bburx=380bp,bbury=685bp}
   \end{center}
   \caption{The orbits of the vortices and antivortices of
Fig.~\ref{fig:head-tail} during their head-tail collision.
The filled circles denote the position of vortices and the open
circles the position of the antivortices at successive and equal
times intervals. The numbers $1,\overline{2},3,\overline{4}$ denote the initial
position of the vortices and antivortices.
   }
   \label{fig:orbit2}
\end{figure}

Here, we consider collisions of two V-A pairs
with different velocities and consequently with different sizes.
The results of our numerical simulations are summarized in
Figs.~\ref{fig:head-on},\ref{fig:head-tail},\ref{fig:orbit1} 
and \ref{fig:orbit2}.
Fig.~\ref{fig:head-on} presents head-on collisions and
Fig.~\ref{fig:head-tail} head-tail collisions
through contour plots of the field $m(x,y)$
at six snapshots during the collision process.
Figs.~\ref{fig:orbit1} and \ref{fig:orbit2} present the corresponding
orbits of the individual vortices
during the interaction process. We have traced the center
of every vortex which was considered to be
at the point where the field $m\!=\!1$.
The result of each process depends essentially
on the difference between the sizes of the two V-A pairs.
For head-on collisions we use $\kappa_1 \!=\! \kappa_4 \!=\! 1,
\kappa_2 \!=\! \kappa_3 \!=\! -1$ and for head-tail collisions
we use $\kappa_1 \!=\! \kappa_3 \!=\! 1, \kappa_2 \!=\! \kappa_4 \!=\! -1$.

\widetext
\narrowtext

Fig.~\ref{fig:head-on}a presents the head-on collision of two pairs
with a small difference between their sizes. We use here the V-A pair
solitons which have been numerically calculated in \cite{semi}.
The initial ansatz in our simulation is the product ansatz of two such pairs.
We have taken the size of the left pair $\Ll\! \simeq\! 4$ 
which corresponds to a velocity $\Vl\!=\!0.27$
and the size of the pair on the right $\Lr \!\simeq\! 6.3$ 
which corresponds to $\Vr\!=\!0.15$. The initial
separation measured on the $x$-axis is $\delta\!=\!25$.
In the scattering process the vortices exchange their partners
and two new identical pairs are formed which are scattered at an angle.
Varying the sizes $\Ll, \Lr$ we observe that 
the angle tends to $90^o$ as the difference in the velocities (and momentums)
of the two pairs is getting smaller. 
This process generalizes the $90^o$ scattering
of two identical solitons \cite{rightangle}.

According to (\ref{eq:linearmom}), the larger soliton (pair on the right)
has also a larger linear momentum.
Conservation of the total momentum
implies that each of the resulting identical 
pairs has a non-zero $x$-component of the
momentum and a velocity to the left.
Fig.~\ref{fig:orbit1}a shows the trajectories followed by each vortex and
antivortex. 
Open circles denote the centers of the antivortices
and filled circles those of the vortices.

We defer for later the case of an intermediate difference between the sizes 
of the two pairs and discuss first the case of a large difference 
(Fig.~\ref{fig:head-on}c and \ref{fig:orbit1}c).
In the latter case one expects that the vortices 
which belong to different pairs will
interact loosely with the vortices of
the other pair. As a result, the two pairs are expected
to travel almost undistracted.
We use the ansatz (\ref{eq:ansatz4}) with the parameters
$\Ll\!=\!4,\;\Lr\!=\!7\Ll\!=\!28$.
The initial separation of the pairs is $\delta\!=\!30$.
The result is close to expectations, that is the small pair passes through
the large one. The distortion in the trajectories should become
smaller as the size of the large pair becomes larger.
This case is thus analogous to soliton interaction in one-dimensional
integrable systems, as has been noted by Aref \cite{aref}.

\begin{figure}
   \begin{center}
   \psfig{file=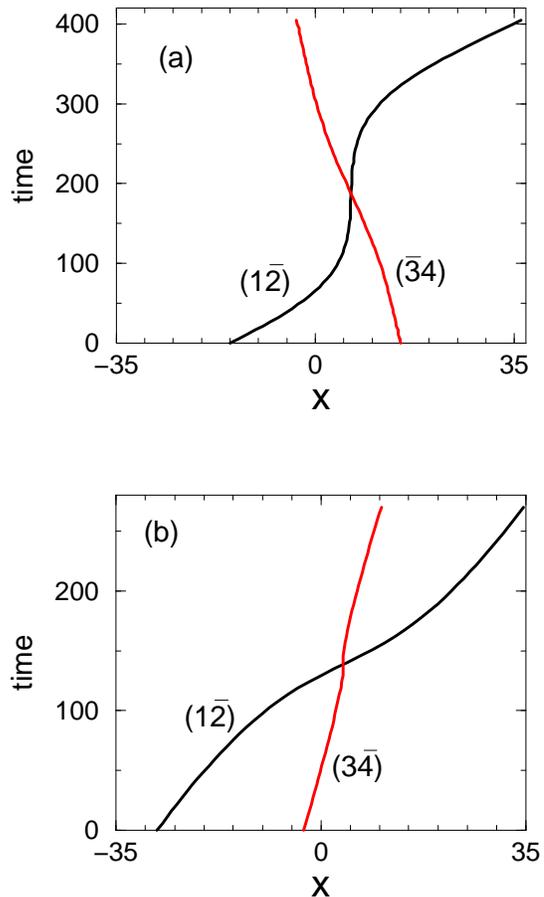,width=7.0cm,bbllx=140bp,bblly=260bp,bburx=410bp,bbury=750bp}
   \end{center}
   \vspace{-10pt}
   \caption{(a) shows the positions on the $x$-axis
of the two pairs $(1\overline{2})$ and
$(\overline{3}4)$ as a function of time, for the simulation
of Figs.~\ref{fig:head-on}c.
(b) shows the positions on the $x$-axis
of the two pairs $(1\overline{2})$ and $(3\overline{4})$
as a function of time, for the simulation
of Figs.~\ref{fig:head-tail}a. See text for details.
   }
   \label{fig:shift}
\end{figure}

In order to explore further this analogy we have plotted in
Fig.~\ref{fig:shift}a the $x$-coordinate of the two pairs
as a function of time. (The data of Fig.~\ref{fig:orbit1}c
correspond to the data of Fig.~\ref{fig:shift}a only until time=345).
We observe that the fast pair experiences a delay during the
interaction with the slow one, which results in a negative
shift (in comparison to the free motion) 
in its $x$-position after the collision. 
On the other hand,
the slow pair is accelerated during the interaction and thus gains 
a positive shift in its $x$-position.
The present observation should be contrasted to the
situation in one-dimensional soliton collisions
where a positive shift for both solitons is observed in all models.

An intermediate situation between those in Figs.~\ref{fig:head-on}a
and \ref{fig:head-on}c is presented in Fig.~\ref{fig:head-on}b. 
The parameters in the initial ansatz are
$\Ll\!=\!4,\; \Lr\!=\!6\Ll\!=\!24,\; \delta\!=\!34$. 
The pairs initially exchange partners
during the scattering process and form new V-A pairs just as in 
the simulation in Fig.~\ref{fig:head-on}a. 
However, after some excursion the  new pairs approach each
other again, exchange partners once more and the initial pairs
re-emerge traveling along their initial direction of motion.
The trajectories followed by the vortices are depicted in
Fig.~\ref{fig:orbit1}b. A similar scenario for point vortices
in hydrodynamics has been discussed in \cite{eckhardt}. 

In the next set of simulations
we explore the situation of a head-tail collision.
That is, both the slow and the fast pair move to the same direction
(to the right in Figs.~\ref{fig:head-tail}a, \ref{fig:orbit2}a).
The parameters here are $\Ll\!=\!8,\; \Lr\!=\!20,\;\delta\!=\!25$.
The two pairs pass through each other, but some differences
to the case of Fig.~\ref{fig:head-on}c should be pointed out.
In Fig.~\ref{fig:shift}b we give the $x$-component of the
trajectories of the pairs. We find that the fast pair is
accelerated during the interaction (positive shift)
while the slow pair is decelerated (negative shift).
(See, however, the relevant remarks in the next section.)

The last simulation, presented in Figs.~\ref{fig:head-tail}b
and ~\ref{fig:orbit2}b, includes two identical V-A pairs
traveling along the same direction. It can be considered as
a limiting case to that of Fig.~\ref{fig:head-tail}a 
when the sizes of the pairs are equal.
The parameter values are $\Ll\!=\!8,\; \Lr\!=\!8,\; \delta\!=\!6$.
The system can be also viewed as a vortex-vortex pair and an 
antivortex-antivortex pair.
Both pairs rotate while at the same time
the magnetic quadrupole which is formed is propagating
along the $x$-axis.
A similar "leap-frogging" motion was studied in hydrodynamics
by Love \cite{love} and has also been observed with
two vortex rings \cite{oshima}.
Since the magnetic quadrupole is characterized by two parameters (the velocity
and the internal frequency) it can be
considered as an analog of a breather \cite{aref,eckhardt}.
We can make an estimate of the mean velocity of propagation
of the quadrupole. Suppose that $L$ is the mean distance of the pair
of vortices from the pair of antivortices. If this is large
compared to the distance between vortices of the same kind we may
consider the quadrupole as two dipoles on top of each other.
Then, a straightforward
generalization of the results of \cite{semi} gives
\begin{equation}
\label{eq:2overl}
v \simeq {2 \over L}, \qquad L\gg 1.
\end{equation}
The rough estimate says
that the quadrupole propagates with twice the velocity of a single
V-A pair. In the present case, we have $L\!=\!8$ which implies a velocity
$v\!\simeq\!0.25$. Indeed, our simulations show that the quadrupole in
Figs.~\ref{fig:head-tail}b, \ref{fig:orbit2}b, 
propagates with a mean velocity $v_q \!\simeq\! 0.24$ while a single V-A pair
with a size $L\!=\!\Ll\!=\!8$ has a mean velocity $v_d \!\simeq\! 0.125$.

The simulations of this section give an overview
The richness of the above
results shows that exploring numerically the different scenarios that occur
during the interaction of two V-A pairs is a rather cumbersome task. 
We note that in the processes that we have been
studying all the vortices retain their identity
during collision. This implies that an
analytical calculation based on collective coordinates should be
successful in reproducing the numerical results and should also
provide an overview of the observed phenomena.
This task will be taken up in the next section. 

\begin{figure}
   \begin{center}
   \psfig{file=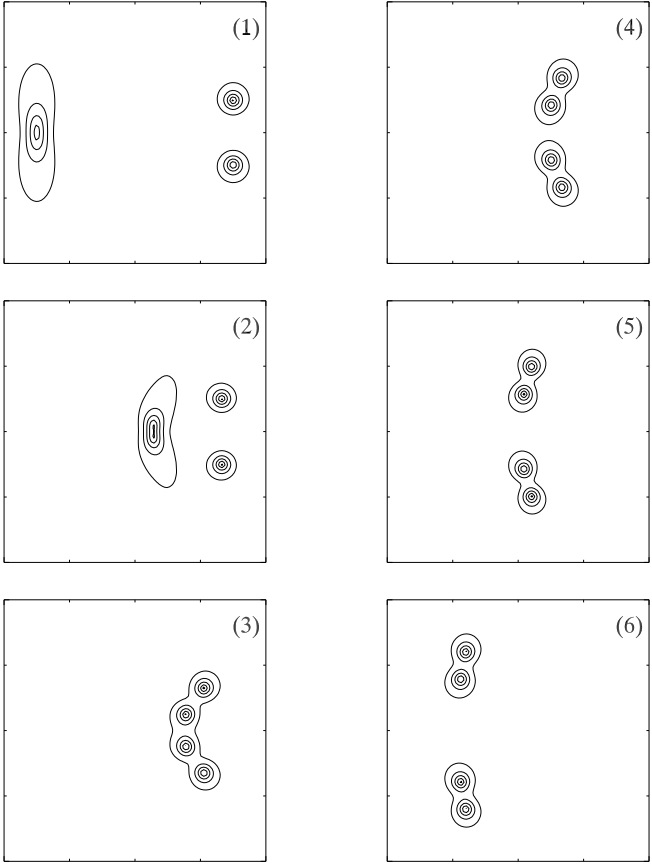,width=6.5cm,bbllx=170bp,bblly=236bp,bburx=390bp,bbury=655bp}
   \end{center}
   \vspace{-20pt}
   \caption{Head-on collision between a slow V-A pair with velocity $\Vr=0.1$
(right in the first entry) and a fast soliton with velocity $\Vl=0.9$
(left in the first entry).
We present six snapshots at times $t=0,22,44,66,88,132$.
The fast soliton is split in a vortex and an antivortex at the time
of collision. The vortices exchange partners and two new pairs
are formed which are scattered at an angle.
   }
   \label{fig:split}
\end{figure}

In the remainder of the present section we continue 
with simulations of collisions of solitons
which do not have a distinct vortex-antivortex pair character.
We use here again the semitopological solitons found in \cite{semi}.
An example is given in Fig.~\ref{fig:split}. In the initial state, 
in the first entry of the figure, the V-A pair on the right
has a velocity $\Vr\!=\!0.1$ while the other soliton on the left has
a large velocity, specifically here $\Vl\!=\!0.9$, and has
no vortex-antivortex character.
The clear numerical result, shown in the remaining five entries of the
figure, is that the fast soliton is split at collision time in a vortex and
an antivortex. The outcome is two identical V-A pairs which are
scattered at an angle.

\begin{figure}
   \begin{center}
   \epsfig{file=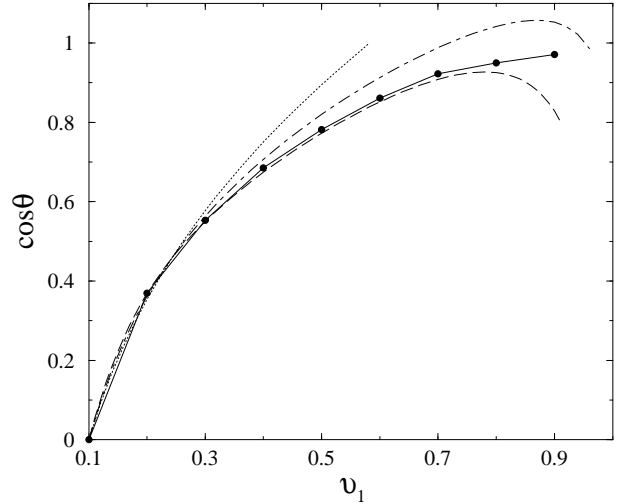,width=8.0cm}
   \end{center}
   \caption{The cosine of the scattering angle for a series of simulations
where the slow soliton has a velocity $\Vr=0.1$ and the fast one
$0.1 \leq \Vl < 1$, is given by the dots which have been connected
by a solid line.
The dashed line results from conservation of energy and linear momentum.
The dotted and the dot-dashed lines give the result of formula
(\ref{eq:angle}) for two different methods of calculating the soliton
lengths (see text for details).
   }
   \label{fig:angle}
\end{figure}

We perform a series of simulations of collisions
between solitons, where one of them has a definite velocity -- we
chose $\Vr\!=\!0.1$ -- while
the velocity of the other varies in the range $0.1\!\leq\!\Vl\!<\!1$.
The results are summarized in Fig.~\ref{fig:angle}
which gives the cosine of the 
scattering angle $\theta$ as a function of the velocity 
$\Vl$ of the fast soliton (filled circles
connected with a solid line in the figure).

\noindent The scattering angle is measured (in Fig.~\ref{fig:split})
from the negative horizontal axis.
We have $\theta\!=\!\pi/2$ when the velocities of the
two solitons are equal, that is when
$\Vl\!=\!\Vr\!=\!0.1$. It is then monotonically decreasing as
$\Vl$ increases until the value $\Vl\!\simeq\!0.9$. 

When the velocity of the second soliton becomes greater than the value
$\Vl\! \simeq \!0.91$
we face a new scenario. The soliton is initially split into a V-A pair,
then follows a loop and finally it re-joins its initial
partner. We eventually obtain the picture of two solitons which have passed
through each other.
The scenario resembles that of Fig.~\ref{fig:head-on}b.
The difference here is that, well after the collision time,
the fast soliton is destabilized.
When the velocity of the fast soliton is increased, the loop that
the vortices follow during collision becomes smaller.
For $\Vl\!\gsim\!0.97$ we have no loop any more, and the fast soliton
passes through the slow pair.

The value $\Vr\!=\!0.1$ used in the simulations in Fig.~\ref{fig:angle}
is not special. We can obtain results similar to those in 
Fig.~\ref{fig:angle} using a different $\Vr$, e.g.
$\Vr\!=\!0.2$ or $0.3$. 

\vspace{10pt}

\noindent The value of $\Vl$ until which
scattering at an angle occurs, seems to be close to the value $\Vl\!=\!0.9$
in these cases, too.
For increasing $\Vr$ the minimum scattering angle, that we can 
obtain by simulations, increases.

The simulations of this section give an overview
of the possible scattering scenarios.
However, the picture will not be complete until we obtain an analytical
understanding. The theory which we present in the next section
accounts for the simulations presented 
in Figs.~\ref{fig:head-on},\ref{fig:head-tail}
and provides a reasonably satisfactory understanding. 
A step towards the understanding of the results of 
Figs.~\ref{fig:split},\ref{fig:angle} will also be taken.

\vbox{\vspace{2cm}}

\section{Collision of Vortex-Antivortex pairs: Analytical description}

A full analytical investigation of interactions between vortex pairs
appears to be quite complicated. Suffice it to say that 
no analytical formula is known for a single vortex pair soliton.
However, one can employ  a rigid shape approximation
and suppose that each vortex is a coherently traveling structure
and is also well separated from all the others.
Then the dynamics of the system of vortices reduces to that
of their centers ${\bf R}_i$. The latter obey the equations
\cite{thiele,huber}
  \begin{equation}
\label{eq:thiele}
{d{\bf R}_i \over dt} \times {\bf G}_i = {\bf F}_i\;,
  \end{equation}
where ${\bf G}_i$ is the gyrocoupling vector (\ref{eq:gyrovector}) 
for vortex $i$
and ${\bf F}_i \equiv - \partial E / \partial {\bf R}_i$. The quantity $E$
is the interaction energy between the vortices, therefore we call
${\bf F}_i$ the force
on vortex $i$ exerted by the other vortices in the system.

To make further progress we need the form of $E$,
that is we need to know the field of the vortices.
Under the assumption that the vortices are "quasi-static"
the field of a single vortex is approximately given by
Eq.~(\ref{eq:vortexphi},\ref{eq:vortexm}).
One should note that the energy is finite only when
  \begin{equation}
\label{eq:kappa}
\sum_i{\kappa_i} = 0,
  \end{equation}
where $\kappa_i$ is the vortex number of a single vortex.
We shall study only this situation here.
Then the energy $E$ of vortex interaction has the form \cite{kosterlitz}
  \begin{equation}
\label{eq:thieleenergy}
E \simeq -2 \pi \; \sum_{i<j} \kappa_i\, \kappa_j\; 
\ln(|{\bf R}_i - {\bf R}_j|),
  \end{equation}
where a constant has been omitted. 

We put ${\bf R}_i\!=\!(X_i,Y_i)$ in Eq.~(\ref{eq:thiele}) and obtain
  \begin{eqnarray}
\label{eq:thielemotion}
{d X_i \over dt} & = & - \sum_{j \neq i}
  \kappa_j\, {(Y_i-Y_j) \over ({\bf R}_i - {\bf R}_j)^2 }, \nonumber \\
\noalign{\medskip}
{d Y_i \over dt} & = & \sum_{j \neq i}
  \kappa_j\, {(X_i-X_j) \over ({\bf R}_i - {\bf R}_j)^2 }.
  \end{eqnarray}
These are the same as the equations of motion of point vortices
in hydrodynamics \cite{lamb,lugt} when the hydrodynamic "vortex strengths",
which correspond to $\kappa_i$ in the present system,
are $\pm 1$.

One can now see that
  \begin{eqnarray}
\label{eq:ormi}
P_x & = & \sum_i \; P_i^{x} = \sum_i  \;2\pi\, \kappa_i\, Y_i\;,\nonumber \\
\noalign{\medskip}
P_y & = & \sum_i \; P_i^{y} = - \sum_i \;2\pi\, \kappa_i\, X_i\;,
  \end{eqnarray}
as well as
  \begin{equation}
\label{eq:stroformi}
\ell = \sum_i \;2\pi\, \kappa_i\, (X_i^2 + Y_i^2)
  \end{equation}
are conserved quantities.
Eqs.~(\ref{eq:ormi}) give the two components of the total momentum
and Eq.~(\ref{eq:stroformi}) gives
the total angular momentum of the system. They can be derived from the
formulas (\ref{eq:linearmom},\ref{eq:angularmom})
when the rigid-shape approximation is used.
We further note that Eqs.~(\ref{eq:thiele}) can be derived as
the Hamilton equations
associated with the Hamiltonian (\ref{eq:thieleenergy}) with the conjugate
variables $X_i$ and $P_i^x \!=\! 2\pi\,\kappa_i\,Y_i$.

In general, system (\ref{eq:thielemotion}) is integrable
only for a system of three point vortices.
However, for a system of two V-A pairs, for which condition (\ref{eq:kappa}) 
is satisfied, Eqs. (\ref{eq:thielemotion}) can be integrated, when the pairs
are initially moving along the same line \cite{eckhardt2}.
This is a fortunate situation since we shall be exclusively concerned
with such systems in the present paper.

The approximations employed so far seem to be quite crude.
For instance we expect the rigid-shape approximation to be valid
when the distances between the vortices are much larger than
the size of the out-of-plane structure of each vortex (unity in our units). 
This could severely restrict the
applicability of Eqs.~(\ref{eq:thielemotion}). However,
the numerical simulations, of the previous section,
indicate that the results of the present approximate theory
could be qualitative correct even for situations
beyond the applicability limits of the theory.

In the following we give the equations of motion of
two V-A pairs. The vortex and antivortex of the first pair
are placed at positions $(\Xl,\pm \Yl)$ and
have $\kappa \!=\! \pm 1$ . We denote this pair schematically as
$(1 \overline{2})$. The second pair is at 
$(\Xr,\pm \Yr)$, has $\kappa \!=\! \mp 1$, and is denoted
$(\overline{3} 4)$. (cf. top entries of Fig.~\ref{fig:head-on}.)
We choose the polarity $\lambda \!=\! 1$ for all vortices.
The system represents two V-A pairs in a head-on collision course.
From Eqs.~(\ref{eq:thielemotion}) we derive the four equations of motion
\begin{eqnarray}
\label{eq:eqmotion1}
{d \Xl \over dt} & = & {1 \over 2 \Yl} + {\Yl-\Yr \over (\Xl-\Xr)^2 + (\Yl-\Yr)^2} 
\nonumber \\
                 & & \qquad  - {\Yl+\Yr \over (\Xl-\Xr)^2 + (\Yl+\Yr)^2},  \\
\noalign{\medskip}
\label{eq:eqmotion2}
{d \Yl \over dt} & = & - {\Xl-\Xr \over (\Xl-\Xr)^2 + (\Yl-\Yr)^2}
  \nonumber \\
                 & & + {\Xl-\Xr \over (\Xl-\Xr)^2 + (\Yl+\Yr)^2},   \\
\noalign{\medskip}
\label{eq:eqmotion3}
{d \Xr \over dt} & = & -{1 \over 2 \Yr} + {\Yl-\Yr \over (\Xl-\Xr)^2 + (\Yl-\Yr)^2}
 \nonumber \\
                 & & \quad \qquad  + {\Yl+\Yr \over (\Xl-\Xr)^2 + (\Yl+\Yr)^2}, \\
\noalign{\medskip}
\label{eq:eqmotion4}
{d \Yr \over dt} & = & -{\Xl-\Xr \over (\Xl-\Xr)^2 + (\Yl-\Yr)^2}
\nonumber \\
                 & & + {\Xl-\Xr \over (\Xl-\Xr)^2 + (\Yl+\Yr)^2}.
\end{eqnarray}
The system (\ref{eq:eqmotion1}-\ref{eq:eqmotion4}) is
completely integrable since there are
four independent conserved quantities. We shall use the energy
and the x-component of the linear momentum:
  \begin{equation}
\label{eq:energeia2}
\Yl \Yr\, {(\Xl-\Xr)^2 + (\Yl-\Yr)^2 \over (\Xl-\Xr)^2 + (\Yl+\Yr)^2} = 
\Ylinit \Yrinit,
  \end{equation}
  \begin{equation}
\label{eq:ormi2}
\Yl- \Yr = \Ylinit - \Yrinit.
  \end{equation}
We suppose that the pairs are initially on the $x$-axis and
at an infinite distance from each other.
Then, $\Ylinit, \Yrinit$ denote the $y$-coordinates 
of the vortices at time $t=-\infty$.

\begin{figure}
   \begin{center}
   \epsfig{file=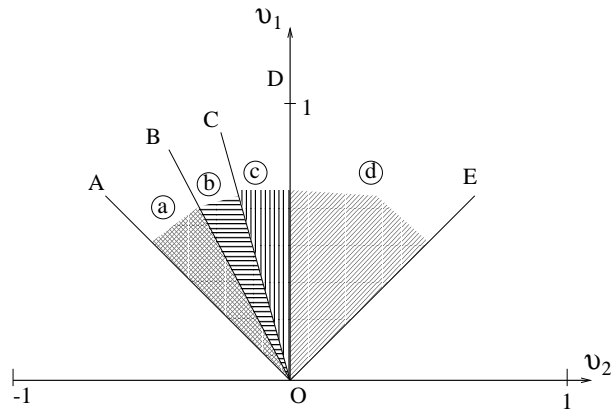,width=8.0cm}
   \end{center}
   \vspace{10pt}
   \caption{A schematic overview of the different possibilities for
head-on and head-tail collision between two soliton pairs.
$\Vl$ and $\Vr$ are the velocities of the two pairs.
The regions a,b and c correspond to to the cases a,b and c of
Fig.~\ref{fig:head-on}. The region d corresponds to the case
of Fig.~\ref{fig:head-tail}a and the scenario of Fig.~\ref{fig:head-tail}b
occurs for velocities on the line OE.
   }
   \label{fig:overview}
\end{figure}

We define $\Ll \!\equiv\! 2 \Ylinit$ and $\Lr \!\equiv\! 2 \Yrinit$
as the sizes of the pairs.
We take $\Lr \!\geq\! \Ll$ while $ \Ll,\Lr \gg 1$.
Therefore we have for the velocities: 
\begin{equation}
\Vl  = {1 \over \Ll}, \qquad
\Vr  = {1 \over \Lr},
\label{eq:velocities}
\end{equation}
that is, $(1 \overline{2})$ is the "small" and fast pair and
$(\overline{3} 4)$ is the "large" and slower pair.
Eqs.~(\ref{eq:eqmotion1}-\ref{eq:eqmotion4}) 
were studied within the framework of
hydrodynamics in \cite{acton,eckhardt,love}.
It was shown that the behavior of the pairs during scattering 
depends on the ratio
 $\alpha \!=\! \Lr/\Ll \!=\! \Vl/\Vr $.
There are the following three cases: 
\begin{itemize}
\item[(a)]{ when 
 $1 \!\leq\! \alpha \!<\! \alpha_1 \!=\! 3\!+\!2\sqrt{2} \!\simeq\! 5.83$,
the pairs change partners during the process and scatter at an angle
(Fig.~\ref{fig:head-on}a).}
\item[(b)]{ In the intermediate case, when $\alpha_1 < \alpha < \alpha_2 \!=\!
(\sqrt{2}+\sqrt{\sqrt{5}-1})/(\sqrt{2}-\sqrt{\sqrt{5}-1}) \!\simeq\! 8.35$,
the vortices change partners during scattering but later they
rejoin their initial partners and travel along the
initial direction of motion (Fig.~\ref{fig:head-on}b).}
\item[(c)]{ When $\alpha \!>\! \alpha_2$
the fast pair passes through the slow one 
(Fig.~\ref{fig:head-on}c).}
\end{itemize}
In Fig.~\ref{fig:overview} we give a schematic representation
of the different regions in the $(\Vl,\Vr )$ plane where
the three scattering processes occur. 

In case (a) the scattering process can be described by
$(1 \overline{2}) + ( \overline{3} 4) \rightarrow (1 \overline{3})
+ (\overline{2} 4)$ which is a schematic representation of the
change of partners during collision.
In the limit of two identical pairs one can use  Love's \cite{love} equation
(\ref{eq:energeia2}) to obtain the following well-known result
\begin{equation}
\label{eq:right}
{1 \over \Xl^2} + {1 \over \Yl^2} = {1 \over (\Ylinit)^2}\,,
\end{equation}
which is in good agreement with the data of the numerical simulation of the
Landau-Lifshitz equation \cite{rightangle}.
In the general case of two different ingoing pairs,
the outgoing pairs will actually be identical. 
Eqs.~(\ref{eq:energeia2},\ref{eq:ormi2}) give
for the size of the two identical outgoing pairs
  \begin{equation}
\label{eq:newsize}
L_{\hbox{out}} = \sqrt{\Ll\, \Lr}.
\end{equation}
This formula is known for point-like vortices in 2D hydrodynamics
of incompressible fluids \cite{lugt}.

The angle of scattering can be found if we use
the laws of conservation of energy and linear momentum:
\begin{equation}
E = {E_1 + E_2 \over 2}, \qquad P\; \cos\theta = {P_2 - P_1 \over 2},
\end{equation}
where $E$ is the energy and $P$ the absolute value of the
linear momentum on the $x$-axis of each of the final pairs. 
The angle $\theta$ is measured from the direction of motion of
the slow pair. 
We obtain
\begin{equation}
\cos\theta = {P_2 - P_1 \over 2 P}.
\label{eq:angle2}
\end{equation}
The calculation of $P$ requires the use of the energy conservation law
and the energy-momentum dispersion relation which is known numerically
\cite{semi}.

Fig.~\ref{fig:angle} shows the numerical results (filled circles connected by
lines) for the $\cos\theta$ as a function
of $\Vl$, keeping a constant $\Vr\!=\!0.1$.
We have $\Vl\!>\!\Vr$.
The dashed line results from Eq.~(\ref{eq:angle2}) and
it is a very good approximation of the simulation results
until $\Vr \sim 0.78$. At this value the dashed line has a maximum.
Until this critical value of $\Vr$ the collision is elastic in the
sense that almost no energy is dissipated in radiation.
We also have to mention that the approximation shown by the dashed line
is getting worse as the velocity $\Vl$ is increasing.

In the approximation of Eqs.~(\ref{eq:eqmotion1}-\ref{eq:eqmotion4})
and (\ref{eq:velocities}) we have for the scattering angle:
  \begin{equation}
\label{eq:angle}
\cos\theta = {\alpha - 1 \over 2 \sqrt{\alpha}} .
  \end{equation}
The limiting cases are $\theta\!=\! \pi/2$ for $\alpha\!=\!1$ and
$\theta\!=\! \pi$ when $\alpha\!=\!\alpha_1$.
Some care is needed in interpreting the angle $\theta\!=\!\pi$ in
the last limiting case. In this case
the new pairs which emerge follow the parabolic orbit
  \begin{equation}
\label{eq:parabola}
\Xl = - {2\, \Yl^2 \over \Lr-\Ll} = - {\Yl^2 \over 2(1+\sqrt{2})\,\Ylinit}.
  \end{equation}
The above results, should be
a good approximation in the limit of large $\Ll,\Lr$.

There is a simple way to apply Eq.(~\ref{eq:angle}) when the two colliding
V-A pairs are slow and the vortex and antivortex in a pair are
well separated. The size of each pair can be taken to be the
distance between
the two points where the spin variable reaches the north pole
and its velocity is the inverse of it.
This method gives fairly good results for simulations with pairs of
the size used in Fig.~\ref{fig:head-on}.

A comparison of Eq.~(\ref{eq:angle}) with numerical results is
given in Fig.~\ref{fig:angle}.
The cosine of the angle $\theta$ for $1\leq \alpha \leq \alpha_1$,
is plotted by a dotted line.
Fig.~\ref{fig:angle} shows $\cos\theta$ as a function
of $\Vl$, for $\Vr\!=\!0.1$. (not as a function of $\alpha$).
We have $\Vl\!>\!\Vr$ and we consider $\alpha\!=\!\Vl/\Vr$.
The dotted line is a poor approximation to the simulation results
for $\alpha\!>\!4\;\; (\Vr\!>\!0.4)$. The
deviation from the numerical points is even qualitatively wrong already for 
$\Vr\gsim 0.583$.

We recall at this point that in our simulations of the previous section
we obtained results which could be understood as interaction processes
between two dipoles even in the cases where the V-A pairs have no apparent
dipole character.
Exploiting this remark we assume that the V-A pair solitons
(even those with large velocities and no apparent vortex-antivortex
character) are dipoles with a length $L$ and a charge $q$.
To be sure, in the limit that the velocity goes to zero,
$L$ should go to the simple definition of length 
described in the previous paragraphs
and $q$ should go to the vortex number $\kappa=\pm 1$.
In general, however, we use a generalization
of  Eqs. (\ref{eq:ormi}) and (\ref{eq:1overl}) 
and write for the momentum and velocity of such a dipole:
  \begin{equation}
\label{eq:dipole}
P = 2\pi\, q L, \qquad
v = {q \over L}.
  \end{equation}
The length and charge of the dipole are then given by
  \begin{equation}
\label{eq:length}
L = \sqrt{P \over 2\pi v}, \qquad q = \sqrt{P v \over 2\pi},
  \end{equation}
where the values for $P, v$ can now be taken from the 
Table of ref. \cite{semi}.
In the limit $v\!\rightarrow\! 0$ we use the relation
$P v\!=\!2\pi$ to find that $L\!=\!1/v\! \rightarrow\! \infty$ and 
$q\! =\! \pm 1$, which is the expected limit.
In the opposite limit $v\!\rightarrow\! 1$, an asymptotic analysis \cite{semi}
gives that $P\! \sim\! 1/\sqrt{1-v^2} \rightarrow \infty$, that is an
infinite length of the dipole and also 
$q\sim 1/\sqrt{1-v^2} \rightarrow \infty$.
For intermediate velocities the length $L$ is finite and it
reaches a minimum for $v\simeq 0.87$. The values for the
charge $q$ are relatively close to unity for $v<0.5$ and
they increase rapidly for $v>0.9$.

The interesting result is presented in Fig.~\ref{fig:angle} by the
dot-dashed line. It is obtained by applying Eq.~(\ref{eq:angle}),
where for $\alpha$ we substitute the ratio of lengths of the
two V-A pairs given from Eq.~(\ref{eq:length}).
The curve is compared quite good with the results of the simulations
even for large velocities. However, after $\Vr=0.9$ the present approach fails
completely. Indeed, we have no reason to believe that
an approximation based on dipoles would be correct in the limit
$v\rightarrow 1$.

We now turn to case (c) where the initial pairs survive throughout
the process. For $\alpha$ large enough an expansion gives,
with an error ${\cal O}(1/\alpha^4)$, the trajectories
\begin{eqnarray}
\label{eq:trajectory}
\Xl & = & \Vl\;t - {4\over \Vl}\, 
   \arctan\left({2 \Vl \Vr \,t }\right), \nonumber \\
\noalign{\medskip}
\Xr & = & - \Vr \;t - \frac{1}{{\alpha}^2} \,\frac{2}{\Vr }\,
\frac{(2\Vl \Vr t)}{1+(2\Vl \Vr t)^2} \nonumber \\
 & & -\frac{1}{{\alpha}^3}\, \frac{4}{\Vr}\,\arctan (2\Vl\Vr t) + ... 
\nonumber \\
\noalign{\medskip}
\Yl & = & \Ylinit \left( 1 + {1 \over \alpha}\;
   {4 \over 1 + (2 \Vl  \Vr \,t)^2} \right), 
  \nonumber \\
\noalign{\medskip}
\Yr & = & \Yrinit \left( 1 + {1 \over \alpha^2}\;
   {4 \over 1 + (2 \Vl  \Vr \,t)^2} \right),
\end{eqnarray}
where $\Vl , \Vr$ are given in Eq. (\ref{eq:velocities})
and the dots in the second equation stand for a lengthy term of order
$1/\alpha^3$ which we shall not need in our analysis.
It follows that, in this approximation, 
the fast pair $(1{\overline 2})$ deviates strongly from the rectilinear motion.
The shifts in the positions of the vortex pairs as time
varies from $-\infty$ to $\infty$, are
$\Delta \Xl \!=\! - 4\pi/\Vl $ for the fast pair and 
$\Delta \Xr \!=\! 4\pi/\Vr \alpha^3$ for the slow one.
(The terms in the dots do not give any shift in the pair position
when times goes from $-\infty$ to $\infty$.)
The shifts are measured relative to the direction of motion of each pair.
The results of the simulations for the vortex system show that 
the fast pair is decelerated during the collision as if it is
repelled by the slow one, in agreement with Eq.~(\ref{eq:trajectory}).
The simulations further show that the slow pair is
accelerated during the collision as if it is attracted by the fast one.
The final result is a positive and negative shift in the
positions of the fast and slow pair, respectively, a phenomenon
which has not been observed in head-on soliton collisions in one dimension.
The obtained trajectories are similar to those for the 2D Euler
equation \cite{overman}.

In case (b), $\alpha_1 \!<\! \alpha \!<\! \alpha_2$, 
the scattering process is represented by the scheme 
$(1 {\overline 2})+({\overline 3} 4) \rightarrow
(1{\overline 3})+({\overline 2}4) \rightarrow
(1{\overline 2})+({\overline 3}4)$.
This means that the vortices exchange partners at a first stage
of  the scattering, then the new pairs follow a looping orbit
and at the final stage the initial partners rejoin and travel along
the initial direction of motion 
(cf. Figs.~\ref{fig:head-on}c, \ref{fig:orbit1}c).
The loop becomes the parabola (\ref{eq:parabola}) in the limit
$\alpha \!=\! \alpha_1$ and it is a cusp when $\alpha \!=\! \alpha_2$.

We now turn to the head-tail collision.
It corresponds to the area denoted (d) in Fig.~\ref{fig:overview}
and the relevant numerical simulations have been given in
Figs.~\ref{fig:head-tail} and \ref{fig:orbit2}.
The case has been studied within a hydrodynamical context in 
\cite{grobli,love,eckhardt}. As Love showed, only a slip-through motion
(Fig.~\ref{fig:head-tail}a) and a leap-frogging motion
(Fig.~\ref{fig:head-tail}b) can occur.

Suppose that vortices 1 and 3 have $\kappa\!=\!1$ while $\overline{2}$ 
and $\overline{4}$ are antivortices and 
have $\kappa\!=\!-1$ (cf. Fig.~\ref{fig:head-tail}a). 
The equations of motion
are modified as follows: the left hand side of (\ref{eq:eqmotion1},
\ref{eq:eqmotion2}) and the first term on the
right hand side of (\ref{eq:eqmotion3}) change their signs.
The conserved quantities (\ref{eq:energeia2},\ref{eq:ormi2}) now read
  \begin{eqnarray}
\label{eq:energeia3}
 & & \Yl\Yr\;{(\Xl-\Xr)^2+(\Yl+\Yr)^2 \over
(\Xl-\Xr)^2+(\Yl-\Yr)^2} = \nonumber \\
 & = & \Ylinit \Yrinit\;
\;{(\Xlinit-\Xrinit)^2+(\Ylinit+\Yrinit)^2 \over
(\Xlinit-\Xrinit)^2+(\Ylinit-\Yrinit)^2}\;,
  \end{eqnarray}
  \begin{equation}
\label{eq:ormi3}
\Yl+\Yr\;=\;\Ylinit+\Yrinit\;.
  \end{equation}
The pairs are initially at
$(\Xlinit, \pm \Ylinit)$ and $(\Xrinit, \pm \Yrinit)$.
Using the above one can find that two pairs which start infinitely far apart 
will pass through each other for any value of $\Ylinit \!<\! \Yrinit$. 

For a large difference
in the size of the pairs $(\alpha \!\gg\! 1)$ we find, by an expansion, the
solution of the equations of motion 
(with an error ${\cal O} \left(\frac{1}{{\alpha}^4}\right)$):
\begin{eqnarray}
\label{eq:trajectory2}
\Xl & = & \Vl \,t + {4 \over \Vl }\; \arctan \;(2\Vl \Vr \,t)\;, 
   \nonumber \\
\noalign{\medskip}
{\Xr} & = & \Vr \,t -\frac{1}{{\alpha}^2} \,\frac{2}{\Vr }\,
   \frac{(2\Vl \Vr t)}{1+(2\Vl \Vr t)^2} \nonumber \\
   & & + \frac{1}{{\alpha}^3}\,\frac{4}{\Vr}\, \arctan (2\Vl\Vr t) + ...
\;, \nonumber \\
\noalign{\medskip}
\Yl & = & \Ylinit\;\left(1-{1 \over \alpha}\;
        {4 \over 1+(2\Vl \Vr \,t)^2}\right)\;, 
 \nonumber \\
\noalign{\medskip}
{\Yr} & = & \Yrinit\;\left(1+{1 \over {\alpha}^2}
\;{4 \over 1+(2\Vl \Vr \,t)^2}\; \right)\;.
\end{eqnarray}
The dots stand for terms of order ${\cal O} \left(\frac{1}{{\alpha}^3}\right)$
which do not contribute to the shift in the position of the pair
when times varies from $-\infty$ to $\infty$.
In the present case, the shift  of the fast pair is to the direction 
of its motion
$\Delta \Xl \simeq 4\pi/\Vl $.
The shift of the slow pair is also positive but small
$\Delta \Xr \simeq 4 \pi/(\Vr\alpha^3)$.
However, the numerical simulations of the previous section 
(Fig.~\ref{fig:shift}b)
have given a small shift for the slow pair opposite to its direction of motion.
We believe that this should be a consequence of the
second term in the second of Eqs. (\ref{eq:trajectory2}) which
is dominant because of our small space and short integration time.
We note here again a difference of the present system with the
situation in 1D. For instance, in KdV the fast soliton acquires 
a positive shift
and the slow one a negative shift due to a head-tail collision.

A difference between the head-on and head-tail collision
for the case of large $\alpha$, is that during the collision the
size of the small pair increases in the head-on collision case
(Eq.~(\ref{eq:trajectory}) and Fig.~\ref{fig:head-on}c) but it decreases
in the head-tail collision case (Eq.~(\ref{eq:trajectory2}) 
and Fig.~\ref{fig:head-tail}a).

In the limit of a small difference of the sizes of the pairs
($\alpha\!\simeq\!1$) the relative shift of the pairs grows and tends to
infinity for identical pairs like
$\Delta\Xl - \Delta\Xr \sim 4/(\Vl -\Vr )$.
In the limit of two identical pairs the asymptotes of the
solution after the scattering have the form
$\Xr \!=\! Vt - \sqrt {t}\;,
\Xl \!=\! Vt + \sqrt {t}\;$ and, at large distances, the distance between
pairs goes as $\Xl-\Xr \!=\! 2 \sqrt{t}$.
 It is interesting to compare this  shift with that for solitons
in 1D systems where
the shift is proportional to the logarithm of the difference
of the velocities of the two solitons ($\Delta \Xl \!\sim \! \ln(\Vl\!-\!\Vr)$)
and the distance between two identical solitons after collision goes like
$\Xl-\Xr \!\sim\! \ln t$.

Two V-A pairs which have
a finite distance between them, will pass through each
other when the following relation holds
   \begin{eqnarray}
   \label{eq:headtailcriterion}
& & {(\Yrinit+\Ylinit)^2\; [(\Yrinit+\Ylinit)^2-2\,(\Yrinit-\Ylinit)^2] \over 
(\Yrinit-\Ylinit)^2} \nonumber \\
\noalign{\medskip}
& & \hbox{\hspace{3cm}}  > (\Xrinit-\Xlinit)^2 \;.
   \end{eqnarray}
In the opposite case, the pairs form a translating 
bound quadrupole state. The translation is accompanied by a rotation 
of the two vortices and the two antivortices around each other.
This leap-frogging motion was first analyzed in \cite{love}.
An example is given in 
Figs.~\ref{fig:head-tail}b and \ref{fig:orbit2}b.
The quadrupole state has lately acquired special interest due
to its relation to breather modes \cite{eckhardt}.

We shall analyze some characteristics of the leap-frogging
motion in the case that the two rotating vortices are well separated
from the two rotating antivortices.
We call $L$ the distance between the two pairs. The two vortices
rotate clock-wise and the two antivortices counter clock-wise.
The translational motion of the quadrupole has the velocity
given in Eq. (\ref{eq:2overl}).
With the further assumption that the distance $\delta$ between the
vortices of the same pair is large compared to the size of a single vortex
(but still small compared to the size of the quadrupole: 
$L,\delta\gg 1, L \gg \delta$),
we find from Eqs. (\ref{eq:trajectory}) the frequency of rotation
\begin{equation}
\label{eq:quadrupolefrequency}
\omega \simeq {2 \over \delta^2} - {2 \over L^2}.
\end{equation}

\section{ Conclusions}

We have presented a detailed study for the head-on and head-tail
collisions between vortex-antivortex pairs in 2D easy-plane ferromagnets.
The V-A pairs are the simplest units which can be found in free translational
motion in our model. However, they have a nontrivial internal structure
which is responsible for their unusual behavior during interaction.
Our study combines numerical simulations, which yield a variety
of interesting scattering scenarios, with a collective variable theory,
which leads to an understanding of the main features of the
scattering scenarios.
The change of partners between V-A pairs during interaction is
the most remarkable effect in our case. It is due to the internal structure 
of the V-A pairs which has been fully taken into account.

In a study of a system of a lot of vortices the mechanisms which have
been described here should play a dominant role in the way to the final
steady state. The relaxation process should include
a multitude of collisions of vortices and the formation
of steady structures.
 
From the point of view of soliton theory in two dimensions, our
results are novel and can be compared with  a variety of studies
in the field. Some comparison is also done with soliton
theory in one space dimension.

The scattering of coherently traveling objects in two dimensions
appears to be of interest in a variety of physical systems.
Objects similar to the V-A pairs studied here exist in systems
in different fields of physics \cite{jones1,jones2,luther,lamb}.
Even in nonequilibrium systems \cite{heilmann,fineberg} studies
similar to the present one have been performed and some similar results
have been derived.

\section*{Acknowledgments}
We thank H. B\"uttner for giving the initial stimulus for this work
and for interesting conversations.
We also thank the authors of \cite{semi} for kindly
providing us the numerical code
for the calculation of the V-A pair solitons.
A.S.K. thanks the University of Bayreuth for its hospitality.
A.S.K. and S.K. acknowledge financial support from the Graduiertenkolleg
''Nichtlineare Spektroskopie und Dynamik''.


\newpage 

\widetext


\begin{thebibliography}{10}

\bibitem{kovalev79} A.S. Kovalev, A.M. Kosevich, and K.V. Maslov,
JETP Lett. {\bf 30}, 321 (1979)
 
\bibitem{kosevich83} A.M. Kosevich, V.P. Voronov, and I.V. Manzhos, 
Sov. Phys. JETP {\bf 57}, 86 (1983)

\bibitem{nikiforov} A.V.  Nikiforov and E.B. Sonin, 
Sov. Phys. JETP {\bf 58}, 373 (1983)

\bibitem{huber}  D.L. Huber, Phys. Rev. B {\bf 26}, 3758 (1982)

\bibitem{gouvea} M.E. Gouvea, G.M. Wysin, A.R. Bishop, and F.G. Mertens, 
Phys. Rev. B {\bf 39}, 11840 (1989)

\bibitem{mertens1}
 F.G. Mertens, G.M. Wysin, A.R. V\"olkel, A.R. Bishop, M.J. Schnitzer,
 in {\sl "Nonlinear Coherent Structures in Physics and Biology"},
 eds. H.H. Spatschek
 and F.G. Mertens, NATO ASI Series, B; Vol. {\bf 329} (Plenum Press,
 New York and London, 1994)

\bibitem{mertens2}
 F.G. Mertens, A.R. Bishop in {\sl "Nonlinear
 Science  at the dawn of the 21st century"} eds. P.L. Christiansen,
 M.P. Soerensen and A.C. Scott, 
 Springer Lecture Notes, (Springer, Berlin, 2000)
 
\bibitem{paptom} N. Papanicolaou and T.N. Tomaras, 
Nuclear Physics B {\bf 360}, 425 (1991)

\bibitem{semi}
N. Papanicolaou and P.N. Spathis, Nonlinearity {\bf 12}, 285 (1999)

\bibitem{cooper}
N.R. Cooper, Phys. Rev. Lett. {\bf 80}, 4554 (1998)

\bibitem{pitaevskii} L.P. Pitaevskii, Sov. Phys. JETP {\bf 13}, 451 (1961)

\bibitem{jones1} C.A. Jones and P.H. Roberts, J. Phys. A: Math. Gen. {\bf 15},
2599 (1982)
\bibitem{jones2} C.A. Jones, S.J. Putterman and P.H. Roberts, J. Phys. A: Math.
Gen. {\bf 19}, 2991 (1986)

\bibitem{luther} B. Luther-Davis, R. Powles, V. Tikhonenko, 
Opt. Lett. {\bf 19}, 1816 (1994)

\bibitem{lamb} H. Lamb, {\sl Hydrodynamics}, 
Cambridge University press (1993)

\bibitem{lugt}
H.J. Lugt, {\sl Introduction to Vortex Theory}, Vortex Flow Press,
Potomac, Maryland (1996)

\bibitem{kosevich90} A.M. Kosevich, B.A. Ivanov, and A.S. Kovalev, 
Phys. Rep. {\bf 194}, 117 (1990)

\bibitem{bogdan} M.M. Bogdan and A.S. Kovalev, 
JETP Letters {\bf 31}, 424 (1980)

\bibitem{afm}
S. Komineas and N. Papanicolaou, Nonlinearity {\bf 11}, 265 (1998)

\bibitem{thiele} A.A. Thiele, Phys. Rev. Lett. {\bf 30}, 230 (1973)

\bibitem{volkel91} A.R. V\"olkel, F.G. Mertens, A.R. Bishop, G.M. Wysin,
Phys. Rev. B {\bf 43}, 5992 (1991)

\bibitem{volkel94} A.R. V\"olkel, G.M. Wysin, F.G. Mertens, A.R. Bishop, and
H.J. Schnitzer, Phys. Rev. B {\bf 50}, 12711 (1994)

\bibitem{papzakr} N. Papanicolaou and W.J. Zakrzewski, Physica D {\bf 80}, 225
(1995)

\bibitem{rightangle}
S. Komineas, Physica D {\bf 155}, 223 (2001)

\bibitem{grobli} W. Gr\"obli, 
{\sl Specielle Probleme \"uber die Bewegung geradliniger
paralleler Wirbelf\"aden}, Z\"urich, Z\"urcher und Furrer (1877)

\bibitem{greenhill}
A.G. Greenhill, Q. J. Math. {\bf 15}, 10 (1878)

\bibitem{aref} H. Aref, Ann. Rev. Fluid. Mech. {\bf 15}, 345 (1983)

\bibitem{eckhardt} B. Eckhardt and H. Aref, 
Phil. Trans. R. Soc. Lond. {\bf A 326}, 655 (1988)

\bibitem{love} A.E.H. Love, Proc. London Math. Soc. {\bf 25}, 185 (1894)

\bibitem{oshima} Y. Oshima, J. Phys. Soc. Jpn. {\bf 45}, 660 (1978)

\bibitem{kosterlitz} J.M. Kosterlitz and D.J. Thouless, J. Phys. C {\bf 6},
1181 (1973)

\bibitem{eckhardt2} B. Eckhardt,  Phys. Fluids {\bf 31}, 2796 (1988)

\bibitem{acton}  E. Acton, J. Fluid Mech. {\bf 76}, 561 (1976)

\bibitem{overman}  E.A. Overman, N.J. Zabusky, J. Fluid. Mech. {\bf 125},
187 (1982)

\bibitem{heilmann}
S. Komineas, F. Heilmann, L. Kramer, Phys. Rev. E {\bf 63}, 11103 (2001)

\bibitem{fineberg}
J. Fineberg, O. Lioubashevski, Physica A {\bf 249} 10 (1998)


\end{thebibliography}
\end{document}